\documentclass{PoS}

\title{Jets from Galactic Binaries}

\ShortTitle{Jets from Galactic Binaries}

\author{\speaker{Thomas J. Maccarone}\\%         
  Faculty of Applied and Physical Sciences\\
        University of Southampton, Southampton - U.K.\\
        E-mail: \email{tjm@soton.ac.uk}}

\abstract{I present a brief review of the properties of jets from
X-ray binaries, highlighting the disk-jet connection, in which there
are strong correlations between X-ray and radio power for black holes
and for neutron star in low/hard spectral states, and reduced emission
in soft states.  I discuss how some of the new ``deviant'' black hole
systems which follow the relation normally found for neutron stars
might fit into such a picture.  I close by highlighting a few open
questions which might be best addressed with soft gamma-ray observations.}

\FullConference{The Extreme and Variable High Energy Sky\\
                 September 19-23, 2011\\
                 Chia Laguna (Cagliari) , Italy}
\begin{document}

\section{Introduction}

Jets are ubiquitous in astrophysics, being seen from nearly all
classes of accreting objects.  In this paper, I will briefly review
the contributions to our understanding of jets that have come from
modelling and observations of Galactic sources, and in the spirit of
this conference's aims, will put a bit of emphasis on where future
hard X-ray and gamma-ray observations may contribute to futher
understanding of jet production and jet physics.

Jets play several important roles in the Universe.  They are one of
the dominant candidates for providing the feedback of energy into the
interstellar medium that stops cooling flows from being found in most
clusters of galaxies.  Jets are also one of the best candidates for
producing the highest energy cosmic rays, which have energies orders
of magnitude higher than those which can be produced in particle
accelerators.  The production of jets may provide key information
about the underlying accretion flows that power them, and the jets
themselves provide important laboratories for testing plasma physics
under extreme conditions.

The viscous timescale from the innermost stable circular orbit in a
Shakura-Sunyaev (1973) accretion disk is about 100 years for a $10^8
M_\odot$ black hole -- and much larger from the outer part of the
disk.  The variability in most AGN thus cannot reflect any fundamental
change in system parameters.  X-ray binaries, on the other hand, can
change in luminosity by 6-7 orders of magnitude on timescales of a
year or so, allowing one to study how the jet power is affected by
accretion rate changes in a single system, so that one does not need
to amass a large sample of objects to control for black hole mass or
spin.  Additionally, X-ray binaries generally are dominated in
bolometric luminosity in the X-rays, meaning that it is
straightforward to understand how the accretion rate is changing
without covering hard-to-observe regions of the spectrum like the
far-infrared or ultraviolet.  These advantages make it much easier to
understand the accretion-ejection connection in X-ray binaries than in
any other class of objects.

\section{Spectral states and the disk-jet connection}

X-ray binaries exhibit a range of ``spectral state'' phenomenology.
These states represent collections of properties of the source, with
sharp changes in the source spectral energy distributions taking place
in concert with sharp changes in the source power spectra.  The
longest lived states are the low/hard state and the high/soft state.
The high/soft state agrees very well with the predictions of the
standard geometrically thin, optically thick, thermally emitting
accretion disk model of Shakura \& Sunyaev (1973).  In the low/hard
state, source spectra are reasonably well fitting by cutoff power laws
with photon indices of about 1.7 and cutoff energies around 100 keV,
well explained by thermal comptonization in a geometrically thick,
optically thin flow (Thorne \& Price 1975).  The sources show strong
red-noise variability with fractional rms amplitude of about 30\%.
Low magnetic field neutron star sources show the same basic
phenomenology, but with slightly more complicated spectra, probably
related to their having both boundary layers and accretion disks, each
of which can contribute quasi-thermal emission.  In both classes of
sources, the transitions are hysteretical, with the transition from
hard to soft usually occurring near the peak of an outburst (Miyamoto
et al 1995; Maccarone \& Coppi 2003), and the transition from
soft-to-hard states usually occurring near 2\% of the Eddington
luminosity (Maccarone 2003).  At the state transitions, hybrid spectra
are seen, which tend to have strong thermal and strong power law
components simultaneously, and the power spectra often show strong
quasi-periodic oscillations.

Two correlations are clearly exhibited between accretion disk
properties and jet power (as measured from the radio flux) in black
hole X-ray binaries.  The first is the switch-off of the radio
emission in the high/soft state (Tananbaum et al. 1972; Fender et
al. 1999).  The standard mechanisms for powering relativistic jets
require large scale height magnetic fields (Livio et al. 1999; Meier
2001), so that the radio jet would be expected to become much less
powerful when the accretion flow becomes geometrically thin.  The
recent finding that the radio power of the persistently soft state
black hole 4U~1957+11 is at least a factor of 300 below the
extrapolation of the hard state relation (Russell et al. 2011) comes
close to presenting a challenge to theory.

The other correlation which has been found is one between X-ray
luminosity and radio luminosity in the hard state (Hannikainen et
al. 1998; Corbel et al. 2000; Gallo, Fender \& Pooley 2003 -- GFP).
Most hard state black hole X-ray binaries have been found to follow a
relation where $L_R \propto L_X^{0.7}$ (GFP).  This relation is
consistent with theoretical expectations under three assumptions: (1)
the kinetic power into the jet scales linearly with the mass accretion
rate onto the black hole (2) the radio power follows standard
synchrotron theory for conical, self-absorbed jets, and hence $L_R
\propto L_{kin}^{1.4}$ and (3) the accretion onto the black hole is
radiatively inefficient, with $L_X \propto \dot{m}^2$, consistent with
expecations from an advection dominated accretion flow (e.g. Narayan
\& Yi 1995).

Two key differences present themselves between the radio properties of
low magnetic field neutron star X-ray binaries and those of black hole
X-ray binaries.  The first is that the radio/X-ray correlation in the
neutron stars seems to follow the $L_R \propto L_X^{1.4}$ relation as
a rule in the neutron stars (e.g. Migliari et al. 2003 -- M03). The
second is that the neutron star systems can show relatively strong
radio emission in soft states (e.g. Migliari et al. 2004).  Both of
these differences can be understood as boundary layer effects.  The
steeper slope for the radio/X-ray relation in the hard state is
straightforward to explain -- the X-ray emission is expected to be
radiatively efficient for accretion onto a neutron star (M03).  The
radio emission in the soft states can be explained as being powered by
the boundary layer itself -- the boundary layer is necessarily
differentially rotation, and necessarily has a scale height comparable
to its distance from the center of the neutron star (e.g. Maccarone
2008).  The magnetic field of the neutron star itself may also
contribute to jet launching (D. Meier, private communication).

In recent years, a few black hole systems have been seen that are
underluminous in the radio relative to the sample studied by GFP.  In
one of the best studied cases, that of H1743-322, Coriat et al. (2011)
showed that the source follows a $L_R \propto L_X^{1.4}$ relation in
the brighter parts of its hard state, before transiting to a more
standard $L_R \propto L_X^{0.7}$ relation deeper into the hard state.
The steeper relation has been seen in several other sources in recent
years.  A possible explanation is that in these ``deviant'' hard
states, the accretion flow is really radiatively efficient.  This
might be explained if, e.g. there exists a geometrically thin,
optically thick accretion disk at the center of the accretion flow --
something that has been suggested in another context by
Meyer-Hofmeister et al. (2009).  Such a disk could be a relic of the
high/soft state, since, due to the high density of gas in the
innermost regions, the evaporation of the thin disk into a thick, hot
disk is expected to happen later in the outburst decay than the
evaporation of the disk at moderate distances from the black hole.
Since the innermost part of the accretion flow would be a thin disk,
one would then expect that such a flow would be radiatively efficient
-- and, in fact, much like the neutron star accretion flows in the
hard state which join to boundary layers in their centers.  It is
worth noting that the radio weak hard states have been seen primarily
in X-ray binaries which fade very slowly from their peak brightnesses,
and that this slow fading may be related to their inability to
evaporate their entire inner disks into hot, thick accretion flows.

\section{Short wavelength emission from jets}
In most cases, in the low hard states, the emission region size in the
radio will be many light minutes across or more, limiting the amount
of rapid variability that can be expected.  As a result, to look for
the most rapid variability from the jet, and hence to probe how it is
powered on short timescales, it is necessary to look at higher
frequency photons, which come from closer in to the black hole or
neutron star.  The development of high-speed solid state photometers
in the optical and infrared have made such measurements possible in
recent years, after decades in which relatively little high speed work
had been done in this wavelength range.  The optical data have been
puzzling, typically showing both a positively correlated component in
the cross-correlation with a moderate time lag, and an anti-correlated
component, often at negative time lag.  This result has been explained
in terms of a magnetic energy reservoir model in which the optical
emission comes from the jet (Malzac et al. 2004), and in terms of a
model in which the optical emission comes from a combination of the
synchrotron emission from the corona whose Compton upscattering
produces the X-rays, and some thermal reprocessing in the outer
accretion disk (Veledina et al. 2011).

At the present time, only one example of sub-second variability in the
infrared has been presented in the literature (Casella et al. 2010).
It is a much simpler observation, with the infrared emission lagging
behind the X-rays by about 0.1 seconds, but with no statistically
significant anti-correlations and with no strong asymmetries to the
cross-correlation function.  Additionally, the brightness temperature
of the infrared emission is sufficiently high that it cannot be
thermal emission without the expectation that the thermal X-ray
emission from that component would exceed the observed X-ray flux.  As
a result, in this case, the infrared emission must really be coming
from the relativistic jet that produced the radio emission, and the
time delay is likely revealing the travel time of perturbations to the
region where the infrared emission is produced, making IR variability
a potentially powerful tool for mapping out how jets are powered.

It is clear that jets may emit detectably at high energies when they
collide with the interstellar medium(e.g. Corbel et al. 2002), or the
stellar winds of their systems' donor stars (e.g. Tavani et al. 2009;
Fermi LAT Collaboration 2009).  On the other hand, the more
interesting question concerns the viability of models which suggest
that the bulk of the hard X-ray emission from ``normal'' low/hard
state systems may come from their jets (e.g. Markoff et al. 2001).

A few pieces of observational evidence indicate that the X-rays do
not, in fact, come from the jet.  In the neutron star X-ray binary
4U~0614+091, the break from an optically thick compact jet to an
optically thin jet can be clearly seen in its broadband spectral
energy distribution, and the extrapolation of the optically thin jet
falls far below the X-ray flux (Migliari et al. 2010).  Additionally,
because jets are radiatively efficient, and cannot carry away too
large a fraction of the total accretion power, one would expect a
sharp drop in the X-ray luminosity at the state change from the
high/soft state to low/hard state if the jet were the dominant source
of X-ray luminosity in the low/hard state, but sharp transitions are
not seen (Maccarone 2005).  Therefore, it seems most likely that the
dominant mechanism for producing the X-rays in the hard states of
X-ray binaries is, in fact, thermal Comptonization, and that the
correlation between X-rays and radio is simply a result of the higher
power to the jet expected at high accretion rate even in the context
of a standard underlying thin or thick accretion disk (e.g. Meier
2001).

\section{Do the black hole spins matter?}

Many recent attempts have been made to measure black hole spins in the
past decade or so.  The two most-used methods are reflection model
fitting (e.g. Ross \& Fabian 2005) and thermal continuum fitting
(e.g. Davis \& Hubeny 2006).  Attempts to correlate either the radio
power, or the ratio of the radio power to the X-ray power with these
spin indicators have not revealed any detectable correlations,
indicating that the black hole spin measurements are in error, or that
the jet power does not correlate strongly with the black hole spin, or
both (Fender, Gallo, Russell 2010).  This result lies in constrast
with some claims from studies of active galactic nuclei, which find
correlations between radio power and properties of their host galaxies
suggested to correlate with black hole merger history, and hence spin
(Balmaverde \& Capetti 2006; Sikora et al. 2007); however, both of
these results neglect to account for the effects of the black hole
mass on the radio flux (e.g. Merloni et al. 2003), and furthermore,
the latter of these papers fails to deal properly with extended
emission which may be resolved out for some objects and not others.
There is thus still no strong observational evidence that black hole
spin is important for powering jets, however strong the theoretical
considerations for it may be.

\section{Some key open questions}

As the purpose of this meeting was partly to serve as the 9th birthday
party for INTEGRAL, I highlight a few open questions in jet physics
that might best be addressed with INTEGRAL.  One is the aforementioned
role of the neutron star's magnetic field, which would be best studied
by getting increased sample sizes of millisecond X-ray pulsars -- a
class of objects INTEGRAL has been especially successful in
discovering.  A second is an understanding of the effects of spectral
state transition hysteresis on jet properties -- again something with
which INTEGRAL can help, since its harder response and better
sensitivity relative to other wide field monitors makes it more
sensitive to early activity of X-ray transients.  An additional one is
the pair composition of jets -- Granat tenatively detected a
redshifted annihilation line from Nova Muscae near state transition
(Gilfanov et al. 1991), but the far more sensitive INTEGRAL still has
not been able to observe an X-ray transient in a similar spectra
state.

\end{document}